\input{epsf}
\documentstyle[12pt]{article}

\def\simge{\mathrel{%
   \rlap{\raise 0.511ex \hbox{$>$}}{\lower 0.511ex \hbox{$\sim$}}}}
\def\simle{\mathrel{
   \rlap{\raise 0.511ex \hbox{$<$}}{\lower 0.511ex \hbox{$\sim$}}}}
 
\def\slashchar#1{\setbox0=\hbox{$#1$}           
   \dimen0=\wd0                                 
   \setbox1=\hbox{/} \dimen1=\wd1               
   \ifdim\dimen0>\dimen1                        
      \rlap{\hbox to \dimen0{\hfil/\hfil}}      
      #1                                        
   \else                                        
      \rlap{\hbox to \dimen1{\hfil$#1$\hfil}}   
      /                                         
   \fi}                                         %
\def\nn{\nonumber}
\def\ts{\thinspace}
\def\tx{\textstyle}
\def\ra{\rightarrow}

\def\ol{\bar}

\def\be{\begin{equation}} 
\def\ee{\end{equation}} 
\def\bea{\begin{eqnarray}}
\def\eea{\end{eqnarray}}
\def\ba{\begin{array}}
\def\ea{\end{array}}
\def\chipr{\chi^{\ts \prime}}
\def\CA{{\cal A}}
\def\CB{{\cal B}}
\def\CC{{\cal C}}
\def\CD{{\cal D}}

\def\CF{{\cal F}}
\def\CG{{\cal G}}

\def\CM{{\cal M}}

\def\CR{{\cal R}}

\def\shat{\hat s}
\def\that{\hat t}
\def\uhat{\hat u}
\def\rshat{\sqrt{\shat}}

\def\atro{\alpha_{\tro}}

\def\Ntc{N_{TC}}

\def\thw{\theta_W}
\def\kslash{\raise.15ex\hbox{/}\kern-.57em k}
\def\tro{\rho_{T}}

\def\tropm{\rho_{T}^\pm}
\def\trop{\rho_{T}^+}
\def\trom{\rho_{T}^-}
\def\troz{\rho_{T}^0}
\def\tom{\omega_T}
\def\tpi{\pi_T}
\def\tpipm{\pi_T^\pm}
\def\tpimp{\pi_T^\mp}
\def\tpip{\pi_T^+}
\def\tpim{\pi_T^-}
\def\tpiz{\pi_T^0}
\def\tpipr{\pi_T^{0 \prime}}

\def\gev{{\rm GeV}}

\def\half{{\textstyle{ { 1\over { 2 } }}}}
\def\third{{\textstyle{ { 1\over { 3 } }}}}
\def\fourth{{\textstyle{ { 1\over { 4 } }}}}

\def\fourthirds{{\textstyle{ { 4\over { 3 } }}}}

\begin{document}
\title{
\vskip -15mm
\begin{flushright}
\vskip -15mm
{\small BUHEP-99-5\\
hep-ph/9903372\\}
\vskip 5mm
\end{flushright}
{\Large{\bf Technihadron Production and Decay Rates in the Technicolor Straw
    Man Model}}\\ 
}
\author{
{\large Kenneth Lane\thanks{lane@buphyc.bu.edu}}\\
{\large Department of Physics, Boston University}\\
{\large 590 Commonwealth Avenue, Boston, MA 02215}\\
}
\maketitle
\begin{abstract}
We present hadron collider production rates for the lightest color-singlet
technihadrons in a simple ``straw-man'' model of low-scale technicolor. These
rates are presented in a way to facilitate their encoding in PYTHIA. This
document is a companion to my paper, ``Technihadron Production and Decay in
Low-Scale Technicolor''.
\end{abstract}


\newpage

\section{The Technicolor Straw Man Model}

In this note we present formulas for the decay rates and production cross
sections for the lightest color-singlet technivector mesons $V_T = \tro$ and
$\tom$. The decay rates have been revised for $V_T \ra G \tpi$, where $G$
is a transversely polarized electroweak gauge boson, $\gamma$, $Z^0$, or
$W^\pm$. The gauge boson polarization is defined relative to the spin
direction of the technivector meson in the latter's rest frame. This is
parallel to the beam direction in a hadron or lepton collider.

Some basic references are: for technicolor and extended
technicolor~\cite{tc,etc}; for walking technicolor~\cite{wtc}; for top
condensate models and topcolor-assisted technicolor
(TC2)~\cite{topcondref,topcref,tctwohill,tctwoklee}; for multiscale
technicolor~\cite{multi}; for signatures of low-scale technicolor in hadron
and lepton colliders~\cite{elw}. As has been emphasized in
Refs.~\cite{multi,elw}, a large number $N_D$ of technifermion doublets are
required in TC2 models with a walking technicolor gauge coupling. In turn,
$N_D$ techni-doublets imply a relatively low technihadron
mass scale, set by the technipion decay constant $F_T \simeq
F_\pi/\sqrt{N_D}$, where $F_\pi = 2^{-1/4} G_F^{-1/2}=246\,\gev$. In the
models of Ref.~\cite{tctwoklee}, for example, the number of electroweak
doublets of technifermions $N_D \simeq 10$ and $F_T \simeq 80\,\gev$.

To set the ground rules for our calculations, we adopt the ``Technicolor
Straw Man Model''. In the TCSM, we assume that we can consider in isolation
the lowest-lying bound states of the lightest technifermion doublet, $(T_U,
T_D)$. These are assumed to be color singlets and to transform under
technicolor $SU(\Ntc)$ as fundamentals; they have electric charges $Q_U$ and
$Q_D$. The bound states in question are vector and pseudoscalar mesons. The
vectors include a spin-one isotriplet $\tro^{\pm,0}$ and an isosinglet
$\tom$. Since techni-isospin is likely to be a good approximate symmetry,
$\tro$ and $\tom$ should be nearly degenerate.~\footnote{Even though $\troz$
and $\tom$ have nearly the same mass, they do not mix much because, as in QCD,
they have rather different decay rates.}

The pseudoscalars, or technipions, also comprise an isotriplet
$\Pi_T^{\pm,0}$ and an isosinglet $\Pi_T^{0 \prime}$. However, these are not
mass eigenstates. In the TCSM, we assume the isovectors are simple
two-state mixtures of the longitudinal weak bosons $W_L^\pm$, $Z_L^0$---the
true Goldstone bosons of dynamical electroweak symmetry breaking in the limit
that the $SU(2) \otimes U(1)$ couplings $g,g'$ vanish---and mass-eigenstate
pseudo-Goldstone technipions $\tpipm, \tpiz$:
\be\label{eq:pistates}
 \vert\Pi_T\rangle = \sin\chi \ts \vert
W_L\rangle + \cos\chi \ts \vert\tpi\rangle\ts.
\ee
Here, $\sin\chi = F_T/F_\pi \ll 1$. Similarly, $\vert\Pi_T^{0 \prime} \rangle
= \cos\chipr \ts \vert\tpipr\rangle\ + \cdots$, where $\chi'$ is another
mixing angle and the ellipsis refer to other technipions needed to eliminate
the technicolor anomaly from the $\Pi_T^{0 \prime}$ chiral current. These
massive technipions are also expected to be nearly degenerate.  However, as
noted in Ref.~\cite{elw}, there may be appreciable $\tpiz$--$\tpipr$
mixing. If that happens, the lightest neutral technipions are ideally-mixed
$\ol T_U T_U$ and $\ol T_D T_D$ bound states.

Technipion decays are induced mainly by extended technicolor (ETC)
interactions which couple them to quarks and leptons~\cite{etc}. These
couplings are Higgs-like, and so technipions are expected to decay into the
heaviest fermion pairs allowed. One exception to this in TC2 is that only a
few GeV of the top-quark's mass is generated by ETC, so there is no great
preference for $\tpi$ to decay to top quarks nor for top quarks to decay into
them. Also, because of anomaly cancellation, the constituents of the
isosinglet technipion $\tpipr$ may include colored technifermions as well as
color-singlets. Then, it decays into a pair of gluons as well as heavy
quarks. Therefore, the decay modes of interest to us are $\tpip \ra c \ol b$
or $c \ol s$ or even $\tau^+ \nu_\tau$; $\tpiz \ra b \ol b$ and, perhaps $c
\ol c$, $\tau^+\tau^-$; and $\tpipr \ra gg$, $b \ol b$, $c \ol c$,
$\tau^+\tau^-$. Branching ratios are estimated from (for the sake of
generality, we quote the energy-dependent widths for technipions of mass
$s^{1/2}$):
\bea\label{eq:tpiwidths}
 \Gamma(\tpi \ra \ol f' f) &=& {1 \over {16\pi F^2_T}}
 \ts N_f \ts p_f \ts C^2_f (m_f + m_{f'})^2 \nn \\ \nn \\
 \Gamma(\tpipr \ra gg) &=& {1 \over {128 \pi^3 F^2_T}} 
 \ts \alpha^2_c \ts C_{\tpi} \ts \Ntc^2 \ts s^{{3\over{2}}} \ts .
\eea
Here, $C_f$ is an ETC-model dependent factor of order one {\it except} that
TC2 suggests $\vert C_t\vert \simle m_b/m_t$; $N_f$ is the number of colors
of fermion~$f$; $p_f$ is the fermion momentum; $\alpha_c$ is the QCD coupling
evaluated at $s^{1/2}$ ($= M_{\tpi}$ for on-shell technipions); and
$C_{\tpi}$ is a Clebsch of order one. The default values of these and other
parameters are tabulated at the end of this note.

\section{$\tro$ Decay Rates}

In the limit that the couplings $g,g' = 0$, the $\tro$ and $\tom$ decay as
\bea\label{eq:vt_decays}
\tro &\ra& \Pi_T \Pi_T = \cos^2 \chi\ts (\tpi\tpi) + 2\sin\chi\ts\cos\chi
\ts (W_L\tpi) + \sin^2 \chi \ts (W_L W_L) \ts; \nn \\\nn\\
\tom &\ra& \Pi_T \Pi_T \Pi_T = \cos^3 \chi \ts (\tpi\tpi\tpi) + \cdots \ts.
\eea
The $\tro$ decay amplitude is
\be\label{eq:rhopipi}
\CM(\tro(q) \ra \pi_A(p_1) \pi_B(p_2)) = g_{\tro} \ts \CC_{AB}
\ts \epsilon(q)\cdot(p_1 - p_2) \ts,
\ee
where, scaling naively from QCD,
\be\label{eq:atrho}
\atro \equiv {g_{\tro}^2\over{4\pi}} = 2.91\left({3\over{\Ntc}}\right)\ts,
\ee
and
\be\label{eq:ccab}
\ba{ll}
\CC_{AB} &= \left\{\ba{ll} \sin^2\chi  & {\rm for} \ts\ts\ts\ts W_L^+ W_L^-
\ts\ts\ts\ts {\rm or} \ts\ts\ts\ts  W_L^\pm Z_L^0 \\
\sin\chi \cos\chi & {\rm for} \ts\ts\ts\ts W_L^+ \tpim, W_L^- \tpip
\ts\ts\ts\ts  {\rm or} \ts\ts\ts\ts W_L^\pm \tpiz, Z_L^0 \tpipm \\
\cos^2\chi & {\rm for} \ts\ts\ts\ts \tpip\tpim  \ts\ts\ts\ts {\rm or}
\ts\ts \ts\ts \tpipm\tpiz \ts.
\ea \right.
\ea
\ee
The energy-dependent decay rates (for $\tro$ mass $\sqrt{\shat}$)
\be\label{eq:trhopipi}
\Gamma(\troz \ra \pi_A^+ \pi_B^-) = \Gamma(\tropm \ra \pi_A^\pm \pi_B^0) =
{2 \atro \CC^2_{AB}\over{3}} \ts {\ts\ts p^3\over {\shat}} \ts,
\ee
where $p = [(\shat - (M_A+M_B)^2) (\shat - (M_A-M_B)^2)]^\half/2\rshat$
is the $\tpi$ momentum in the $\tro$ rest frame.

For $g,g' \ne 0$, $\tro$ decay to transversely polarized electroweak bosons
plus a technipion, $G\tpi$ with $g=\gamma,Z^0,W^\pm$, and to
fermion-antifermion pairs, $f \ol f'$ with $f,f' = q$ or $\ell^\pm,
\nu_\ell$. The decay rate for $V_T \ra G \tpi$ is~\cite{tcsm}
\be\label{eq:trogtpi}
\Gamma(\tro \ra G \tpi) = {\alpha V^2_{\tro G\tpi} \ts p^3\over {3M_V^2}} +
{\alpha A^2_{\tro G\tpi} \ts p\ts(3 M_G^2 + 2p^2)\over {6M_A^2}} \ts,
\ee
where $M_G$ is the G-boson's mass and $p$ its momentum; $M_V$ and $M_A$ are
mass parameters of order several hundred~GeV. The quantities $V_{V_T G\tpi}$
and $A_{V_T G\tpi}$ are defined for $V = \tro, \tom$ as follows:~\footnote{We
have neglected decays such as $\troz \ra W_T W_L$ and $\troz \ra W_T
W_T$. The rate for the former is suppressed by $\tan^2 \chi$ relative to the
rate for $\troz \ra W_T \tpi$ while the latter's rate is suppressed by
$\alpha$.}
\be\label{eq:VA}
V_{V_T G\tpi} = {\rm Tr}\biggl(Q_{V_T} \{Q^\dagger_{G_V}, \ts
Q^\dagger_{\tpi}\}\biggr) \ts,\qquad
A_{V_T G\tpi} = {\rm Tr}\biggl(Q_{V_T} [Q^\dagger_{G_A}, \ts
Q^\dagger_{\tpi}]\biggr) \ts.
\ee
In the TCSM, with electric charges $Q_U$, $Q_D$ for $T_U$, $T_D$, the
generators $Q$ in Eq.~(\ref{eq:VA}) are given by
\bea\label{eq:charges}
Q_{\troz} &=& {1\over{\sqrt{2}}} \left(\ba{cc} 1 & 0 \\ 0 & -1 \ea\right)
\ts;\qquad
Q_{\trop} = Q^\dagger_{\trom} = \left(\ba{cc} 0 & 1 \\ 0 & 0 \ea\right)\nn\\
Q_{\tpiz} &=& {\cos\chi\over{\sqrt{2}}} \left(\ba{cc} 1 & 0 \\ 0 & -1 \ea\right)
\ts;\ts\quad
Q_{\tpip} = Q^\dagger_{\tpim} = \cos\chi\left(\ba{cc} 0 & 1 \\ 0 & 0
  \ea\right)\nn\\
Q_{\tpipr} &=& {\cos\chi'\over{\sqrt{2}}} \left(\ba{cc} 1 & 0 \\ 0
  & 1 \ea\right) \nn\\
Q_{\gamma_V} &=& \left(\ba{cc} Q_U& 0 \\ 0 & Q_D \ea\right)
\ts;\qquad
Q_{\gamma_A} = 0 \nn\\
Q_{Z_V} &=& {1\over{\sin\thw \cos\thw}} \left(\ba{cc} \fourth - Q_U
  \sin^2\thw & 0 \\ 0 & -\fourth - Q_D \sin^2\thw \ea\right) \nn\\
Q_{Z_A} &=& {1\over{\sin\thw \cos\thw}} \left(\ba{cc} -\fourth & 0 \\ 0
  &  \fourth \ea\right) \nn\\
Q_{W^+_V} &=& Q^\dagger_{W^-_V} = -Q_{W^+_A} = -Q^\dagger_{W^-_A} = {1\over
  {2\sqrt{2}\sin\thw}}\left(\ba{cc} 0 & 1 \\ 0 & 0 \ea\right)
\eea
The $V_{V_T G\tpi}$ and $A_{V_T G\tpi}$ are listed in Table~1 below.

{\begin{table}{
\begin{tabular}{|c|c|c|}
\hline
Process & 
$V_{V_T G\tpi}$ & 
$A_{V_T G\tpi}$  
\\
\hline\hline
$\tom \ra \gamma \tpiz$& $c_\chi$ & 0 \\
$\ts\ts\ts\quad \ra \gamma \tpipr$ & $(Q_U + Q_D)\ts c_{\chipr}$ & 0 \\ 
$\qquad \ra Z^0 \tpiz$ & $c_\chi\cot 2\thw$ & 0 \\ 
$\ts\qquad \ra Z^0 \tpipr$ & $-(Q_U+Q_D)\ts c_{\chipr}\tan \thw$ & 0 \\ 
$\ts\ts\ts\ts\qquad \ra W^\pm \tpimp$ & $c_\chi/(2\sin\thw)$ & 0 \\ 
\hline
$\troz \ra \gamma \tpiz$ & $(Q_U + Q_D)\ts c_\chi$ & 0 \\
$\ts\ts\ts\quad \ra \gamma \tpipr$ & $c_{\chipr}$ & 0 \\
$\qquad \ra Z^0 \tpiz$ & $-(Q_U+Q_D)\ts c_\chi \tan \thw$ & 0 \\
$\ts\qquad \ra Z^0 \tpipr$ & $c_{\chipr}\ts \cot 2\thw$ & 0 \\
$\ts\ts\ts\ts\qquad \ra W^\pm \tpimp$ & 0 & $-c_\chi/(2\sin\thw)$ \\
\hline
$\tropm \ra \gamma \tpipm$ & $(Q_U + Q_D)\ts c_\chi$ & 0 \\ 
$\qquad \ra Z^0 \tpipm$ & $-(Q_U+Q_D)\ts c_\chi \tan \thw$ & $c_\chi
\ts /\sin 2\thw$ \\  
$\ts\ts\ts\qquad \ra W^\pm \tpiz$ & 0 & $c_\chi/(2\sin\thw)$ \\ 
$\ts\ts\ts\qquad \ra W^\pm \tpipr$ & $c_{\chipr}/(2\sin\thw)$ & 0 \\
\hline\hline
\end{tabular}}
\caption{Amplitudes for $V_T \ra G \tpi$ for $V_T = \tro,\tom$ and $G$ a
  transverse electroweak boson, $\gamma,Z^0,W^\pm$. Here, $c_\chi = \cos\chi$
  and $c_{\chipr} = \cos\chipr$.}
\end{table}}

The $\tro$ decay rates to fermions with $N_f=1$ or 3~colors
are~\footnote{Eqs.~(\ref{eq:trhoff}), (\ref{eq:afactors}) and
(\ref{eq:tomff}) below correct Eqs.~(3) and (6) in the second paper and
Eqs.~(3) and (5) in the third paper of Ref.~\cite{elw}. A factor of
$M^4_{V_T}/\shat^2$ that appears in Eqs.~(6) and~(11) of that second paper
has been eliminated from Eqs.~(\ref{eq:trhoff}) and~(\ref{eq:tomff}). This
convention is consistent with the off-diagonal $s f_{G V_T}$ terms in the
propagator matrices $\Delta_{0,\pm}$ defined in Eqs.~(\ref{eq:gzprop})
and~(\ref{eq:wprop}) below. For weakly-coupled narrow resonances such as
$\tro$ and $\tom$, the difference is numerically insignificant.}
\bea\label{eq:trhoff}
\Gamma(\troz \ra f_i \ol f_i) &=& {N_f \ts \alpha^2 p\over
{3 \atro \shat}} \ts \left((\shat - m_i^2)
\ts A_i^0(\shat) + 6 m_i^2\ts \CR e(\CA_{iL}(\shat)
\CA_{iR}^*(\shat))\right) \ts, \nn\\ \\
\Gamma(\trop \ra f_i \ol f'_i) &=& {N_f \ts \alpha^2 p\over
{6 \atro} \shat^2} \ts \left(2\shat^2 - \shat (m_i^2 + m^{'2}_i) -
(m_i^2 - m^{'2}_i)^2\right) A_i^+(\shat) \ts,\nn
\eea

where I assumed a unit CKM matrix in the second equality. The quantities
$A_i$ are given by
\bea\label{eq:afactors}
A_i^\pm(\shat) &=& {1 \over {8\sin^4\thw}} \ts \biggl\vert{\shat \over {\shat
    - \CM^2_W}}\biggr\vert^2 \ts, \nn \\ \nn\\
A_i^0(\shat) &=& \vert \CA_{iL}(\shat) \vert^2
+ \vert \CA_{iR}(\shat) \vert^2 \ts,
\eea
where, for $\lambda = L,R$,
\bea\label{eq:zfactors}
\CA_{i\lambda}(\shat) &=& Q_i + {2 \zeta_{i\lambda} \ts \cot 2\thw \over
   {\sin 2\thw}} \ts \biggl({\shat \over {\shat - \CM^2_Z}}\biggr)\ts, \nn\\
\zeta_{i L} &=& T_{3i} - Q_i \sin^2\thw \ts, \nn\\
\zeta_{i R} &=& - Q_i \sin^2\thw \ts.
\eea
Here, $Q_i$ and $T_{3i} = \pm 1/2$ are the electric charge and left-handed
weak isospin of fermion $f_i$. Also, $\CM^2_{W,Z} = M^2_{W,Z} - i \rshat \ts
\Gamma_{W,Z}(\shat)$, where $\Gamma_{W,Z}(\shat)$ is the weak boson's
energy-dependent width.~\footnote{Note, for example, that $\Gamma_Z(\shat)$
includes a $t\ol t$ contribution when $\shat > 4m_t^2$.}.

\section{$\tom$ Decay Rates}

We assume that the 3-body decays of $\tom$ to technipions, including
longitudinal weak bosons, are kinematically forbidden. This leaves 2-body
decays to technipions, $G\tpi$ and $f_i \ol f_i$.

The rates for the isospin-violating decays $\tom \ra \pi^+_A \pi^-_B =
W^+_L W^-_L$, $W^\pm_L \tpimp$, $\tpip \tpim$ are given by
\be\label{eq:isov}
\Gamma(\tom \ra \pi^+_A \pi^-_B) = \vert\epsilon_{\rho\omega}\vert^2 \ts
\Gamma(\troz \ra \pi^+_A \pi^-_B) =
 {\vert\epsilon_{\rho\omega}\vert^2 \atro \CC^2_{AB} \over{3}} \ts {\ts\ts
   p_{AB}^3\over {s}} \ts,
\ee
where $\epsilon_{\rho\omega}$ is the isospin-violating $\tro$-$\tom$ mixing
amplitude. In QCD, $\vert \epsilon_{\rho\omega}\vert \simeq 5\%$, so we
expect this decay mode to be negligible if this is chosen to be the nominal
value of this parameter.

The $\tom \ra G\tpi$ decay rates involving a transversely polarized
electroweak boson $G = \gamma,Z,W$ have the same form as
Eq.~(\ref{eq:trogtpi}):
\be\label{eq:tomgtpi}
\Gamma(\tom\ra G \tpi) = {\alpha V^2_{\tom G\tpi} \ts p^3\over {3M_V^2}} +
{\alpha A^2_{\tom G\tpi} \ts p\ts(3 M_G^2 + 2p^2)\over {6M_A^2}} \ts,
\ee

The $\tom$ decay rates to fermions with $N_f$ colors are given by
\be\label{eq:tomff}
\Gamma(\omega_T \ra \ol f_i f_i) = {N_f \ts \alpha^2 p\over
{3 \atro \shat}} \ts \left((\shat - m_i^2)
\ts B_i^0(\shat) + 6 m_i^2\ts \CR e(\CB_{iL}(\shat)
\CB_{iR}^*(\shat))\right) \ts, 
\ee
where
\bea\label{eq:bfactors}
&B_i^0(\shat) &= \vert \CB_{iL}(\shat) \vert^2
+ \vert \CB_{iR}(\shat) \vert^2 \ts, \nn \\ \nn\\
&\CB_{i\lambda}(\shat) &= \left[Q_i - {4 \zeta_{i\lambda} \sin^2\thw \over
    {\sin^2 2\thw}} \biggl({\shat \over {\shat - \CM_Z^2}}\biggr)\right] \ts
(Q_U + Q_D)
\ts.
\eea

\section{Cross Sections for $q_i \ol q_j \ra \tro,\tom \ra X$}

The subprocess cross sections presented below assume that initial-state
quarks are massless. All cross sections are averaged over the spins and
colors of these quarks. These cross sections require the propagator matrices
in the neutral and charged spin-one channels, $\Delta_0$ and $\Delta_\pm$.
The $\gamma$--$Z^0$--$\troz$--$\tom$ propagator matrix is the inverse of
\be\label{eq:gzprop}
\Delta_0^{-1}(s) =\left(\ba{cccc}
s & 0 & -s f_{\gamma\tro} & -s f_{\gamma\tom} \\
0 & s - \CM^2_Z  & -s f_{Z\tro} & -s f_{Z\tom} \\
-s f_{\gamma\tro}  & -s f_{Z\tro}  & s - \CM^2_{\troz} & 0 \\
-s f_{\gamma\tom}  & -s f_{Z\tom}  & 0 & s - \CM^2_{\tom} 
\ea\right) \ts.
\ee
Here, $f_{\gamma\tro} = \xi$, $f_{\gamma\tom} = \xi \ts (Q_U + Q_D)$,
$f_{Z\tro} = \xi \ts \cot 2\thw$, and $f_{Z\tom} = - \xi \ts
(Q_U + Q_D) \tan\thw$, where $\xi = \sqrt{\alpha/\atro}$.
The $W^\pm$--$\tropm$ matrix is the inverse of
\be\label{eq:wprop}
\Delta_{\pm}^{-1}(s) =\left(\ba{cc} s - \CM^2_W & -s f_{W\tro} \\ -s
  f_{W\tro} & s - \CM^2_{\tropm} \ea\right) \ts,
\ee
where $f_{W\tro} = \xi/(2\sin\thw)$.

The rates for production of any technipion pair, $\pi_A\pi_B = W_L W_L$,
$W_L\tpi$, and $\tpi\tpi$, in the isovector ($\tro$) channel are:
\bea\label{eq:pippim}
& &{d\hat\sigma(q_i \ol q_i \ra \troz \ra \pi^+_A\pi^-_B)
  \over{d\that}} = \nn\\
& & \qquad{\pi \alpha\atro \CC^2_{AB} 
  (4\shat p^2 -(\that-\uhat)^2) \over{12 \shat^2}} \ts
\biggl(\vert\CF^{\tro}_{iL}(\shat)\vert^2 +
  \vert\CF^{\tro}_{iR}(\shat)\vert^2\biggr)\ts.
\eea
and
\be\label{eq:pippiz}
{d\hat\sigma(u_i \ol d_i \ra \trop \ra \pi^+_A\pi^0_B)
  \over{d\that}} = 
{\pi \alpha\atro \CC^2_{AB} 
  (4\shat p^2 -(\that-\uhat)^2) \over{24 \sin^2\thw \shat^2}} \ts
\vert\Delta_{W\tro}(\shat)\vert^2 \ts.
\ee
where $p = [(\shat - (M_A+M_B)^2) (\shat - (M_A-M_B)^2)]^\half/2\rshat$ is the
$\shat$-dependendent momentum of $\pi_{A,B}$. As usual, $\that = M^2_A -
\rshat(E_A - p\cos\theta)$, $\uhat = M^2_A - \rshat(E_A + p\cos\theta)$,
where $\theta$ is the c.m. production angle of $\pi_A$.  The factor $4\shat
p^2 -(\that-\uhat)^2 = 4 \shat p^2 \sin^2\theta$. The quantities
$\CF^{V_T}_{i\lambda}$ for $\lambda = L,R$ in Eq.~(\ref{eq:pippim}) are
given in terms of elements of $\Delta_0$ by
\be\label{eq:ffactors}
\CF^{V_T}_{i\lambda}(\shat) = Q_i \ts \Delta_{\gamma V_T}(\shat)
  + {2 \zeta_{i\lambda} \over{\sin 2\thw}} \Delta_{Z V_T}(\shat) \ts.
\ee
Because the $\tro$-$\tom$ mixing parameter $\epsilon_{\rho\omega}$ is
expected to be very small, the rates for $q_i \ol q_i \ra \tom \ra \pi^+_A
\pi^-_B$ are ignored here.

The cross section for $G\tpi$ production in the neutral channel is
given by
\bea\label{eq:gpineutral}
&& {d\hat\sigma(q_i \ol q_i \ra \troz,\ts \tom \ra G \tpi)
\over{d\that}} = \nn \\
&& \quad {\pi \alpha^2 \over{24 \shat}} \Biggl\{
\left(\vert\CG^{VG\tpi}_{iL}(\shat)\vert^2 +
      \vert\CG^{VG\tpi}_{iR}(\shat)\vert^2\right) \ts
      \left({\that^2 + \uhat^2 -2M^2_G M^2_{\tpi}\over{M^2_V}}\right) 
\\
&& \qquad + \left(\vert\CG^{AG\tpi}_{iL}(\shat)\vert^2 +
             \vert\CG^{AG\tpi}_{iR}(\shat)\vert^2\right) \ts
\left({\that^2 + \uhat^2 -2M^2_G M^2_{\tpi} + 4\shat
    M^2_G\over{M^2_A}}\right)\Biggr\} \ts, \nn
\eea
where, for $X = V,A$ and $\lambda = L,R$,
\be\label{eq:gfactors}
\CG^{XG\tpi} _{i\lambda} = \sum_{V_T = \troz,\tom} X_{V_T G \tpi} \CF^{V_T}
_{i\lambda} \ts.
\ee
The factor $\that^2 + \uhat^2 -2M^2_G M^2_{\tpi} = 2\shat p^2
(1+\cos^2\theta)$. The $G\tpi$ cross section in the charged channel is given
by (in the approximation of a unit CKM matrix)
\bea\label{eq:gpicharged}
& &{d\hat\sigma(u_i \ol d_i \ra \trop \ra G \tpi) \over{d\that}} =
{\pi \alpha^2 \over{48 \sin^2\thw \ts \shat}} \ts \vert\Delta_{\ts
  W\tro}(\shat)\vert^2 \\
& & \times \left\{{V^2_{\trop G\tpi}\over{M^2_V}}\biggl(\that^2 + \uhat^2
  -2M^2_G  M^2_{\tpi}\biggr) +
{A^2_{\trop G\tpi}\over{M^2_A}}\biggl(\that^2 + \uhat^2 -2M^2_G
M^2_{\tpi} + 4\shat M^2_G\biggr)\right\} \ts.\nn
\eea

The cross section for $q_i \ol q_i \ra f_j \ol f_j$ (with $m_{q_i} =
0$ and allowing $m_{f_j} \ne 0$ for $t \ol t$ production) is
\bea\label{eq:qqffrate}
{d\hat\sigma(q_i \ol q_i \ra \gamma ,\ts Z \ra \ol f_j f_j)
  \over{d\that}} &=& 
{N_f \pi \alpha^2\over{3\shat^2}} \biggl\{\left((\uhat-m_{f_j}^2)^2 +
  m_{f_j}^2\shat\right)
\ts \left(\vert\CD_{ijLL}\vert^2 + \vert\CD_{ijRR}\vert^2\right) \nn\\
&+& \left((\that-m_{f_j}^2)^2 +
  m_{f_j}^2\shat\right)\ts\left(\vert\CD_{ijLR}\vert^2 + 
  \vert\CD_{ijRL}\vert^2\right)\biggr\} \ts, 
\eea
where
\bea\label{eq:dfactors}
\CD_{ij\lambda\lambda'}(\shat) &=& Q_i Q_j \ts
\Delta_{\gamma\gamma}(\shat)  + {4\over{\sin^2 2\thw}} \ts \zeta_{i \lambda}
\ts \zeta_{\j \lambda'} \ts \Delta_{ZZ}(\shat) \\
&& + {2\over{\sin 2\thw}} \ts \biggl(\zeta_{i \lambda} Q_j
\Delta_{Z\gamma}(\shat) + Q_i \zeta_{j \lambda'} \Delta_{\gamma
  Z}(\shat)\biggr)
\ts. \nn 
\eea
Finally, the rate for the subprocess $u_i \ol d_i \ra f_j \ol f'_j$ is
\be\label{eq:udffrate}
{d\hat\sigma(u_i \ol d_i \ra W^+ \ra  f_j \ol f'_j)
  \over{d\that}} = {N_f \pi \alpha^2\over{12\sin^4\thw \ts \shat^2}} \ts
(\uhat - m^2_j)(\uhat-m^{'2}_j) \ts \vert\Delta_{WW}(\shat)\vert^2 \ts.
\ee

\section{Default Values for Parameters}

The suggested default values of the parameters used in this note are listed
in Table~2.

{\begin{table}{
\begin{tabular}{|c|c|}
\hline
Parameter&  Default Value
\\
\hline\hline
$\Ntc$& 4 \\
$\sin\chi$  & $\third$ \\
$\sin\chi'$  & $\third$ \\
$Q_U$  & $\fourthirds$ \\
$Q_D = Q_U-1$  & $\third$ \\
$C_b$& 1\\
$C_c$& 1\\
$C_\tau$& 1\\
$C_t$& $m_b/m_t$\\
$C_{\tpi}$& $\tx{4\over{3}}$\\
$\vert\epsilon_{\rho\omega}\vert$ & 0.05 \\
\hline
$F_T = F_\pi \sin\chi$  & $82\,\gev$ \\
$M_{\tropm}$ & $210\,\gev$ \\
$M_{\troz}$ & $210\,\gev$ \\
$M_{\tom}$ & $210\,\gev$ \\
$M_{\tpipm}$ & $110\,\gev$ \\
$M_{\tpiz}$ & $110\,\gev$ \\
$M_{\tpipr}$ & $110\,\gev$ \\
$M_{V}$ & $200\,\gev$ \\
$M_{A}$ & $200\,\gev$ \\

\hline\hline
\end{tabular}}
\caption{Default values for parameters in the Technicolor Straw Man Model.}
\end{table}}

\vfil\eject

\end{document}